\documentclass[a4paper,11pt]{article}
\pdfoutput=1 

\usepackage{jheppub} 

\usepackage[T1]{fontenc} 

\newcommand{\be}{\begin{equation}}
\newcommand{\ee}{\end{equation}}
\newcommand{\beq}{\begin{eqnarray}}
\newcommand{\eeq}{\end{eqnarray}}

\title{Effects of Non-Conformal Boundary on Entanglement Entropy}

\author[a]{Andrew Loveridge,\note{Corresponding author.}}

\affiliation[a]{University of Texas at Austin,\\2515 Speedway, Austin, Texas 78712, USA}

\emailAdd{aloveridge@utexas.edu}

\abstract{Spacetime boundaries with canonical Neuman or Dirichlet conditions preserve conformal invarience, but ``mixed'' boundary conditions which interpolate linearly between them can break conformal symmetry and generate interesting Renormalization Group flows even when a theory is free,
providing soluble models with nontrivial scale dependence. We compute the (Rindler) entanglement entropy for a free scalar field with mixed boundary conditions in half Minkowski space and in Anti-de Sitter space.
In the latter case we also compute an additional geometric contribution, which according to a recent proposal then collectively give the 1/N corrections to the entanglement entropy of the conformal field theory dual.
We obtain some perturbatively exact results in both cases which illustrate monotonic interpolation between ultraviolet and infrared fixed points.
This is consistent with recent work on the irreversibility of renormalization group, allowing some assessment of the aforementioned proposal for holographic entanglement entropy and illustrating the generalization of the g-theorem for boundary conformal field theory.}

\begin{document} 
\maketitle
\flushbottom

\section{Introduction} \label{SI}

The entropy of entanglement, famously elucidated in early debates on the nature of quantum mechanics\footnote{
The phrase ``entropy of entanglement'' is modern terminology \cite{Srednicki:1993im} but the physics was understood}
, has emerged more recently as a powerful
unifying tool in quantum field theory. It can act as an order parameter for phase transitions \cite{Kitaev:2005dm} \cite{b} \cite{Amico:2007ag} \cite{Hertzberg:2010uv}, has provocative links with the black hole entropy formula \cite{tHooft:1984kcu} \cite{Solodukhin:2011gn},
has assisted in generalizing the $c$-theorem to higher dimensions \cite{Casini:2012ei} \cite{Casini:2017vbe}, and appears to play a role in the holographic emergence of spacetime geometry \cite{Ryu:2006bv} \cite{Lin:2014hva} \cite{Dong:2016eik} in the $AdS/CFT$ correspondence 
\cite{Maldacena:1997re} \cite{Witten:1998qj} \cite{Aharony:1999ti}.

Despite the utility of the quantity, it remains notoriously difficult to compute. Most examples involve only free fields \cite{Casini:2009sr}
 or exploit conformal symmetry \cite{Holzhey:1994we} \cite{Calabrese:2009qy}.
 This is unfortunate since some highly interesting applications involve the renormalization group flow of the entanglement entropy,
as alluded to above.

\begin{figure}[h!]
\label{EntropyBoundary}
\includegraphics[width=\textwidth]{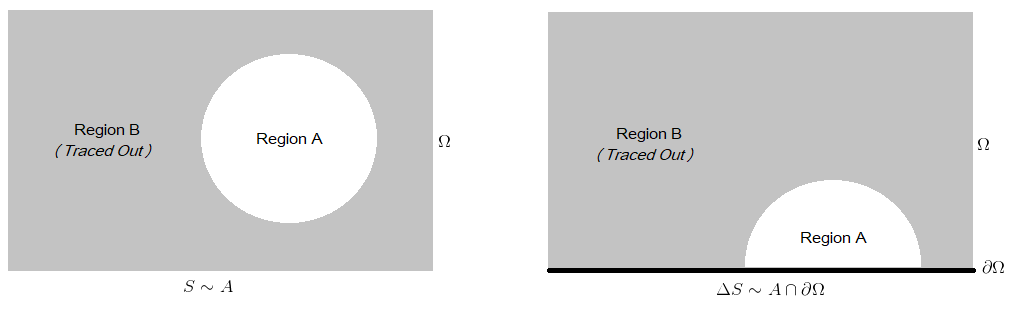}
\centering
\caption{The entanglement entropy of Region A in a spacetime without boundary is proportional to its surface area, $S \sim A$. If one introduces
a boundary with associated boundary physics, then the difference in entanglement entropy for difference choices of boundary physics scales
with the intersection of the region with the boundary, $\Delta S \sim \partial A = A \cap \partial \Omega$}.
\end{figure}

For a spacetime without boundary the leading contribution
 to the entanglment entropy of a quantum field theory is the \emph{area law} \cite{Srednicki:1993im}. 
That is, if one considers the entropy as a function of the scale of 
the region, the dominant contribution is proportional to the surface area. If one introduces a spacetime boundary, it is possible to have
an additional term which scales as the intersection of the surface area with the boundary (this may or may not be subleading depending on the
geometry of the spacetime). The difference in entanglement entropy for different choices of boundary physics $\Delta S$ will appear in this term. See
Figure \ref{EntropyBoundary}.

\begin{figure}[h!]
\label{BulkBoundaryFig}
\includegraphics[width=\textwidth]{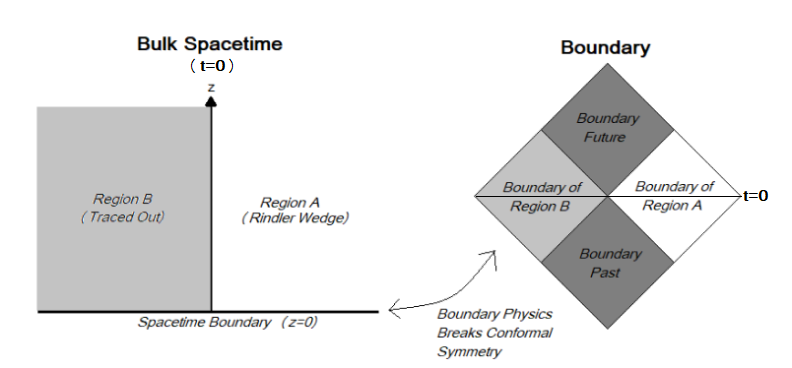}
\centering
\caption{We consider bulk spacetimes with boundary, either half Minkowski space or Anti-de Sitter spaces. Both have boundaries which are 
Minkowski space of one lower dimension. Mixed boundary conditions break conformal symmetry of the boundary,
 where for the latter we are thinking holographically.
We are interested in computing the entanglement entropy of a Rindler wedge whose horizon intersects
the $z$ axis.}
\end{figure}

In this paper we will explore two simple examples in this vein which sidestep the difficulty  with broken
scale invariance mentioned above by imposing ``mixed'' boundary
conditions\footnote{
It is common in the $AdS/CFT$ literature to use this terminology, and we will use it here. However, traditionally ``mixed'' refers to imposing 
different boundary conditions at different \emph{locations} on the boundary whereas we have in mind the linear mixture of boundary terms
traditionally called ``Robin'' boundary conditions
}
 on a free scalar field. By ``mixed" we mean a boundary condition which is some linear interpolation between two canonical conformally
invariant boundary conditions, controlled by a parameter $f$. See Figure \ref{BulkBoundaryFig}.
In particular, we will be interested in the cases where the bulk geometry is Minkowski space or Anti-de Sitter space. For Minkowski space we will include
an artificial boundary at $z=0$ so we are working in half Minkowski space. Anti-de Sitter space already has a (conformal) boundary region.
In both cases the bulk theory is free and the interesting (conformal symmetry breaking) physics is located at the boundary, which is a $d$ dimensional
Minkowski space where where $D=d+1$ is the bulk spacetime dimension. The case of Anti-de Sitter space is especially interesting since the bulk space
is holographically dual to a theory associated with boundary.

In either case one may implement the boundary condition via the addition of a boundary action, so one may think of the imposition of mixed boundary
conditions as an insertion of a (relevant) boundary operator of dimension $\Delta=d-[f]$ ($ < D$). In the Minkowski case,
one may think of this as a mass term localized on the boundary. This generates a
renormalization group flow complete with ultraviolet and infrared fixed points which are the conformally invariant theories.
But because the field remains free, the physics is determined entirely by the Green's function and so the entanglement entropy 
(of the Rindler wedge) is exactly computable with the usual methods augmented by some standard tools 
from asymptotic analysis.

The main results are the expressions for the half Minkowski Rindler entropy (\ref{Result1}) and the dual conformal field theory Rindler entropy
(\ref{Result2}). They will conveniently take the form (Exactly in the Minkowski case, to leading order in Anti-de Sitter):
\be
\Delta S^f= \beta(f \epsilon^{d-\Delta}) \Delta S^{\infty}
\ee 
Where $\beta$ is some function which  interpolates between $0$ and $1$ as $f$ interpolates between $0$ and $\infty$.
Also see the corresponding graphs in Figure  \ref{MinkowskiDS} and Figure \ref{plotAdS} which illusrate this behavior of the entropy as
a function of the boundary coupling $f$. The expression (\ref{Result2b}) is also interesting as a distinct, but ultimately subleading, contribution
(with the same behavior).
The results are interesting for three reasons.

First, it is useful to have exact results for the entanglement entropy that are not conformal field theories and which capture the full renormalization
group interpolation between ultraviolet and infrared fixed points. (see \cite{Berthiere:2016ott} for a related example). We find that the interpolation
is monotonic in the parameter $f$. In half Minkowski space, this extends the results of \cite{Berthiere:2016ott} to the case of
 $m^2=0$ where conformal symmetry
is broken \emph{only} by boundary physics.
 In Anti-de Sitter space this extends the results of \cite{Miyagawa:2015sql} and \cite{Sugishita:2016iel} to finite $f$.

Second, the results therefore illustrate the scaling behavior of the entangement entropy, which can be compared with expectations based on the  irreversability of the renormalization group, as follows.
As already mentioned, it has long been known that the dominant contribution to the entanglement entropy\footnote{
As long as the theory in question is considered at zero temperature and not a topological quantum field theory}
obeys an area law \cite{Srednicki:1993im}:
\be
S_{EE} = \mu(g, \epsilon) \frac{A}{\epsilon^{D-2}}
\ee
where $A$ is the \emph{surface area} of the region in question and $\mu$ is some dimensionless parameter which is a function of the cutoff scale 
$\epsilon$ and the coupling constants $g$. Note this is divergent. We may understand this term, and its divergence, as emerging from the short
distance correlations across the boundary (e.g. in the two point fucntion) which persist to arbitrarily short distances.

For a
conformal field theory in $D=2$ this $\mu(g,\epsilon)$ is just a constant proportional to the central charge, as was found in \cite{Holzhey:1994we}
Heuristically this makes sense as the central charge is related to the number of degrees of freedom, which we might expect to scale with correlations
across the boundary. Given the role of $c$ in establishing the irreversability of the renormalization group \cite{Zamolodchikov:1986gt},
this is already a hint of the way
entanglement entropy may serve as a probe of renormalization group flow. Indeed, it was proven in \cite{Casini:2016udt} that the area term decreases monotonically
along the renormalization group flow, which is equivalent to the $c$-theorem in $D=2$

For a spacetime with boundary, one can have an additional term:
\be
\gamma(g, \epsilon) \frac{\partial A}{\epsilon^{d-2}}
\ee
Where $\partial A$ is the area of the intersection of surface area with the boundary. It may or may not be subleading depending on the bulk geometry.
It is this term that concerns us. Other subleading terms are possible with or without a boundary. One may collect all terms and define:
\be
\label{gammaterm}
\tilde{\mu}(g, \epsilon, r)=\frac{S_{EE}'(r)}{(D-2)r^{(D-3)}}
\ee

Where $S_{EE}(r)$ is the entanglement entropy for a spherical region of radius $r$.
It was shown in \cite{Casini:2017vbe} that Strong Subaddivity \cite{SSA}
 along with Lorentz invarience and the ``Markov property'' \cite{Casini:2017roe}
 which applies to the vacuum of a quantum field theory, implies that (among other things)
the renormalization group flow of the entanglement entropy must obey:
\be
\label{SSA_Req}
\Delta \tilde{\mu}_{IR}  \le  \Delta \tilde{\mu}_{UV}
\ee
Where the $\Delta$ here refers to subtracting the corresponding entanglement entropy of the pure (unperturbed) ultraviolet theory and 
``infrared'' means $r \to \infty$ while ``ultraviolet'' means $r \to 0$.\footnote{It's worth noting that neither side is necessarily positive definite}
Hueristically, the point is that this generalized area law coefficient $\tilde{\mu}$ is a quantity which depends on scale, 
 interpolating between ultraviolet and infrared fixed points, and this behavior is constrained by the irreversability of
the renormalization group flow.

Strictly speaking, for half Minkowski space the boundary breaks the bulk Lorentz Invarience,
 so the result (\ref{SSA_Req}) does not apply. However the monotonicity of our
result is illustrative 
\footnote{The result \cite{Casini:2018nym} was actually proven while this work was in the process of publication! Building on the work in \cite{Fursaev:2016inw}}
of the entropic $g$-theorem
\footnote{The $g$-theorem was originally proposed in a non-entropic context, much like the $c$-theorem, in \cite{Affleck:1991tk}, was proven using the boundary beta-function in \cite{Friedan:2003yc}, and extended to $D=3$ in \cite{Jensen:2015swa} }
\cite{Casini:2016fgb} found for boundary conformal field theories in $D=2$
and which was generalized in \cite{Casini:2018nym}, which applies specifically to the second term (\ref{gammaterm}).
This confirms the conventional wisdom that results such as (\ref{SSA_Req}) reflect, again more fundamentally, the
irreversability of the renormalization group which we might expect to apply for more general backgrounds and appear in whatever way is appropriate
given the nature of the physics involved in conformal symmetry breaking.

Third, the results for Anti-de Sitter space, when combined with a geometric contribution which we also compute,
 provide a test for the recent proposal \cite{Faulkner:2013ana} for the $\frac{1}{N}$ corrections
to the celebrated Ryu-Takayangi formula \cite{Ryu:2006bv}. The full statement is that the entanglement entropy of the conformal field theory dual to the bulk Anti-de Sitter
space, which we think of as embedded in a holographic quantum theory of gravity,
is given by:
\be
\label{FLMp}
S_{EE}^{CFT, \partial \gamma}=\frac{A_{\gamma}}{4G}+\frac{\delta A_{\gamma}}{4G}+S_{EE}^{AdS, \gamma}+S_{Wald}+S_{ren}
\ee
The first term ($\sim N^2$) is the Ryu-Takayanagi term, the other terms ($\sim N^0$) include the 1-loop correction to the area, the bulk entanglement
entropy, and corrections from curvature couplings and renormalization counterterms.

In the context of $AdS/CFT$, the mixed boundary conditions have an interesting interpretation as dual to double trace operators in the conformal
field theory, as was first pointed out in \cite{Witten:2001ua}
 and was elaborated on in e.g. \cite{Gubser:2002zh}. In particular, the $f$ quantity is dual to a \emph{coupling constant}
for a double trace interaction:
\be
\sim \frac{f}{2} \int dx^d \hat{\mathcal{O}} \hat{\mathcal{O}}
\ee
Where $\hat{\mathcal{O}}$ is the operator dual to the bulk field in the conformal case $f=0$. This is a very rich topic, which e.g. includes interesting
relations with the stability and boundedness \cite{Troost:2003ig} \cite{Casper:2017gcw} \cite{Cottrell:2017gkb}
of the operators in $AdS/CFT$. Since the entropy may act as an order parameter for phase transitions,
it too probes these issues and we will discuss them, though that will not be our main focus here.

The important point is that 
the result (\ref{SSA_Req}) \emph{does} of course
apply to the conformal field theory dual, and we will attempt to assess whether the prediction of (\ref{FLMp}) obeys the inequality
as expected for any choice of $f$ for which the corresponding operator is relevant. This extends the work of \cite{Miyagawa:2015sql}
 and \cite{Sugishita:2016iel}, which partly inspired this work \cite{Faulkner:2013ana}.

As a side note, we would also 
like to point out that this appears to serve in general as a tractable example of a holographic renormalization group flow generated
entirely by $\frac{1}{N}$ effects, and also that we are able to use our methods to improve the computation of the vacuum energy found in 
\cite{Gubser:2002zh} \footnote{We'll actually compute the Free Energy Density in the conformal field theory}.

\section{Preliminaries and Methods} \label{SP}

The entropy of entanglement is defined as the von Neuman entropy of the
 ``reduced'' density operator associated with some subsector of the full quantum theory.
 Traditionally, one imagines separating the Hilbert space into a direct product:
\be
\label{ABsplit}
H=H_{A} \otimes H_B
\ee
And then taking the trace of the density operator $\rho_{AB}$ for the full state over,
 say, space B to get a ``reduced'' density operator and associated entanglement entropy:
\be
\rho_A=Tr_B \rho_{AB}
\ee
\be
S_{EE}^A=-Tr_A(\rho_A \ln(\rho_A))
\ee
Strictly speaking, for quantum field theories the splitting in (\ref{ABsplit}) is not possible due to the Reeh-Schlieder theorem \cite{RS1},
 and one must instead define the entanglement entropy for a \emph{subring of observables} rather than a subsector of the Hilbert space.
 This distinction turns out to be important for e.g. gauge fields \cite{Casini:2013rba} \cite{Ohmori:2014eia} \cite{Radicevic:2016tlt}
, but since we are here interested only in scalar fields we may ignore this.\footnote{
In fact this subtlety produces ``boundary effects'' for entanglement of its own sort, which are \emph{different} than those investigated here}.
 In quantum physics $S_{EE}^A$ can be nonzero even if the full state $\rho_{AB}$ is pure,
 which is arguably one of the more profound differences from classical theory.

In this work, we are interested in the entanglement entropy associated specifically with the \emph{Rindler Wedge},
 the subregion of spacetime accessible to a uniformally accelerating observer.
 In the case of half Minkowski space, we are imagining a Rindler wedge associated with an
 observer accelerating away from the horizon but remaining equidistant from the artificial boundary.
 In the case of Anti-de Sitter space we are actually imagining a Rindler observer in the \emph{dual conformal field theory}.
 See again figure (\ref{BulkBoundaryFig}). It's worth noting here that the entanglement entropy should be associated not with a spacial slice
 but with the entire causal diamond which is the causal development of the slice.
 The Penrose diagram of the boundary in (\ref{BulkBoundaryFig}) shows the diamond associated with the bulk spacial slice.

We will compute the entanglement entropy using the replica method. In general, this means observing that:
\be
S=-\partial_n |_{n=1}Tr(\rho^n)
\ee
This is just a mathematical fact, but it can be heuristically interpreted as saying the entropy is a measure of the decrease in the
 ``coincidence probability'' (the probability that all systems are found in the same state) as the number of systems is increased.
 Specifically in quantum field theory, this method geometrizes nicely because the $\rho^n$ can be thought of as
 ``gluing'' together multiple copies of the space and then the tracing procedure just computes the partition function on this space:
\be
\label{Seq1}
S=-\partial_n|_{n=1}(Z_n/Z_1^n)=\partial_n|_{n=1}(W_n-n W_1)
\ee
Where $W$ refers to the connected function (or free energy in the Euclidean picture).
 In the case of the Rindler Wedge, the replica manifold is the cone with surplus angle $\theta= 2 \pi (n-1)$ and the conical singularity is located
 at the horizon (which is a single point after we Euclidienize). Within this computational scheme the $\epsilon$ cutoff serves to regulate this singularity,
but the origin of the divergence is still better understood from the heuristic description given in the previous section.

This quantity is divergent in quantum field theory due to the short distance behavior of the Green's function
 which encodes correlations across the horizon at all scales.
 We will therefore introduce a short distance cutoff $\epsilon$ to regulate the result.
 The ``area law'' result mentioned in the introduction implies that the Rindler entropy is also infrared divergent since the Rindler horizon area is infinite,
 so we will include a long distance cutoff $\Lambda$ where necessary as well.

In a free theory, even with nontrivial boundary conditions, the whole theory is determined entirely by its Green's function,
 which may be determined by solving the equation of motion or built from the spectrum of the theory. For example we have that, at one loop:
\be
\label{Seq2}
W=\frac{1}{2}\int_{m^2}^{\infty}dm^2Tr(G)
\ee
Where here the trace is over the spacetime. We may obtain results for the replica manifold from those for $n=1$ by means of 
the Sommerfeld Formula \cite{Sommerfeld}:
\be
\label{Seq3}
F_{2 \pi \alpha}(z)=F_{2 \pi}(z)-\frac{1}{4 \pi i \alpha}\int_{\Gamma} \cot \Big( \frac{w-z}{2 \alpha} \Big)F(w) dw
\ee
Where the $\Gamma$ contour goes down the line $-\pi$ and back up the line $\pi$ and where $F_{2 \pi}(z)$ is an arbitrary $2 \pi-$periodic function

In effect then, computing the entanglement entropy is just a matter of composing the linear functions (\ref{Seq1}, \ref{Seq2}, \ref{Seq3}).
 So the entanglement entropy up to one loop is a linear functional of the Green's function,
 which is expected given the theory is free and we are interested in vacuum correlations across the horizon.
 We will actually be interested here only in the \emph{entropy difference} for different boundary conditions,
 so this linearity is convenient since it means the
 \emph{difference in entanglement entropies is just a linear functional of the difference in Green's functions}.

For Anti-de Sitter space we will need to additionally include a geometric contribution.
 This will be found using the linearized Einstein equation and by point splitting the Green's function to obtain the stress tensor, see Section (\ref{SAdS}).
 This too is linear in the Green's function.

For all cases we will define the subtracted entropy as:
\be
\label{entropydiff}
\Delta S^f=S^f-S^0
\ee
Where $f$ is a boundary coupling which breaks conformal invariance.
 In general $\Delta$ will be used for this subtraction while $\delta$ will be used for quantum corrections,
 although $\Delta$ will also be used for the scaling dimension of some operators where it is not too unclear to do so (we hope).

We will be working with a free massive quantum scalar field. We can define the theory by specifying the (Euclidien)\footnote{
to be clear, $t=-i\tau$} action and partition function:
\be
\label{ScalarFieldTheory}
Z=\int [d \phi]_{\partial \Omega}e^{-I[\phi]} \qquad I[\phi]=\frac{1}{2} \int_{\Omega} \sqrt{g} \big( g^{ab}\partial_a \phi \partial_b \phi+m^2 \phi^2 \big)
\ee
Where $\Omega$ is some spacetime background and the restriction on the path integral must be chosen to implement suitable boundary conditions,
 of which there will be a 1 parameter family. The requirement is that the operator associated with the equation of motion resulting from the action:
\be
\label{Dop}
\hat{D}=- \nabla^2+m^2= -\frac{1}{\sqrt{g}} \partial_a \big(\sqrt{g}g^{ab}\partial_b \quad \big)
\ee
Is a positive operator on the space of functions on satisfying said boundary conditions. This is made manifest if one recognizes that we may integrate by parts to schematically obtain: 
\be
Z=\int [d \phi]_{\partial\Omega} e^{- \phi^{\dagger} \cdot \hat{D} \cdot \phi}
\ee
We will use the variables $D=d+1=n+3$ so that $D$ represents the bulk spacetime dimension, $d$ the boundary dimension, and $n$ the dimension of intersection of the horizon with the boundary.

\section{half Minkowski Space}

\label{SM}

The (Euclidien) background geometry for half Minkowski space is given by:
\be
ds^2=dz^2+d \tau^2+d \vec{x}\cdot d \vec{x} \qquad z \geq 0
\ee
We will imagine a horizon at, say, $x_1=0$, cutting the bulk spacetime in two and allowing us to compute an associated entanglement entropy.

The differential operator which defines the scalar field theory reduces from (\ref{Dop}) to: 
\be
\hat{D}=-\partial_z^2-\vec{\partial_x} \cdot \vec{\partial_x}+m^2
\ee
Meanwhile one may integrate the action by parts to obtain that the operator is positive on the space of functions which obey: 
\be
\partial_z \phi |_{z=0}=f \phi|_{z=0} \qquad f\geq0
\ee
For some fixed $f$. The case $f=0$ is traditionally called the ``Neumann'' Theory while $f \to \infty$ is the ``Dirichlet'' theory. We may think of nonzero $f$ as inserting a relevant boundary operator of dimension $D-2$.

We will be interested in the spectrum of this operator since this can be used to define the quantum field theory and in particular may give us
 the Green's function. So we seek to solve: 
\be
\hat{D} \phi=\lambda \phi
\ee
One may check that the following functions are an orthonomal set of eigenfunctions which obey the boundary conditions: 
\be
\label{eigenfunctionsM}
\phi=\frac{1}{(2 \pi)^{D/2}}(\alpha_{\kappa}\psi_{\kappa}+\alpha_{\kappa}^* \psi_{\kappa}^*)e^{i \vec{k} \cdot \vec{x}}
\ee
Where: 
\be
\psi_{\kappa}=e^{i \kappa z} \qquad \alpha_{\kappa}=\frac{1}{\sqrt{2}} \frac{\kappa+i f}{\sqrt{f^2+\kappa^2}}
\ee
The corresponding Eigenvalue is: 
\be
\lambda= \kappa^2+|\vec{k}|^2+m^2
\ee
Notice that the f=0 case returns: 
\be
\alpha_{\kappa}=\alpha^*_{\kappa}
\ee
Whereas $f \to \infty$ gives:
\be
\alpha_{\kappa}=-\alpha^*_{\kappa}
\ee 
Which are precisely the results we would expect (symmetry and antisymmtry) using the method of images to obtain the spectrum
 with boundary conditions from the spectrum on the whole space. Indeed, one can see from the structure of the eigenfunctions that it is a
 wavelength-dependent generalization of the method images, where the phase of the image depends on the scale.
 The boundary condition induces a sort of ``RG Flow'' from a free scalar field with a ``Neuman mirror'' in the UV (at large $\kappa$) to
 the same theory but with a ``Dirichlet Mirror'' in the IR (small $\kappa$).\footnote{To
 be clear, since the phase depends on scale, an object emitting a range of wavelengths would not really pervieve this as a mirror}

We may find the Green's function from the spectral theory of $\hat{D}$ using the relation: 
\be
\hat{D}^{-1}=\sum_{\lambda}\frac{v_{\lambda}^\dagger v_{\lambda}}{\lambda}
\ee
Where the $v_{\lambda}$ are the eigenfunctions of $\hat{D}$, in this case (\ref{eigenfunctionsM})
We have:
\be
G^{f}=G+\int \frac{d \kappa dk^d}{(2 \pi)^D} \frac{\kappa^2-f^2}{\kappa^2+f^2}  \frac{e^{i\vec{k}\cdot(\vec{x}-\vec{x}')}e^{i\kappa(z+z')}}{\kappa^2+|\vec{k}|^2+m^2}
\ee
Where G is the usual Minkowski Green's function. We see that we have:
\be
G^0=G+\hat{P}_zG \qquad G^{\infty}=G-\hat{P}_zG \qquad \hat{P}_z\equiv z' \to -z'
\ee
We are actually most interested in the ``subtracted'' Green's function since we want to compute differences between the theories: 
\be
\label{dGm}
\Delta G^{f} \equiv G^f-G^0=-2 \int \frac{d \kappa dk^d}{(2 \pi)^D} \frac{f^2}{\kappa^2+f^2}  \frac{e^{i\vec{k}\cdot(\vec{x}-\vec{x}')}e^{i\kappa(z+z')}}{\kappa^2+|\vec{k}|^2+m^2}
\ee
We may now use the formulas from (\ref{SP}) to compute the entanglement entropy.
 We will be cursory in the following, see Appendix A for more details.
 We will choose the case $m^2=0$ since then it is only the boundary that breaks conformal invairence, and $D=4$ for simplicity,
 but our method is generalizable.

Let's start by considering the limiting cases $f \to 0, \infty$ We find:
\be
\label{Minkfinty}
\Delta S^{\infty} = -\frac{1}{24 \sqrt{\pi}} \frac{\Lambda}{\epsilon} \qquad \Delta S^{0} =0
\ee
Where the latter is true by definition. These are the endpoints, we'd like to be able to see the full interpolation as a function of $f$.
 We may try to proceed by expanding the integral (\ref{dGm}), which is intractable, in either small or large $f$:
\be
\frac{f^2/k^2}{1+f^2/\kappa^2}=\sum_n(-f^2/\kappa^2)^{n+1} \qquad f\ll 1
\ee
\be
\frac{1}{1+\kappa^2/f^2}=\sum_n(-\kappa^2/f^2)^n \qquad f\gg 1
\ee
It was pointed out in \cite{Berthiere:2016ott} that the entropy is not an analytic function of $f$ at $f=0$, posssibly due to the appearance of
a tachyon for $f < 0$ which indicates the onset of a phase transition at this point. Therefore we will expand in large $f$.
We obtain:
\be
\Delta S^f = -\frac{1}{24 \sqrt{\pi}} \frac{\Lambda}{\epsilon}\sum_{n=0}^{\infty} \frac{\Gamma(\frac{1}{2}+n)}{\sqrt{\pi}(1+2n)} \Big(\frac{-1}{f^2 \epsilon^2}\Big)^n
\ee
This is divergent \footnote{Consider e.g. the ratio test}, but can be interpreted as an asymptotic series.\footnote{
A theorem of analysis \cite{AAAMiller} ensures that given the integrand is analytic and that the integration is finite term by term, the expression
is the correct asymptotic series}
 Indeed, (\ref{dGm}) is not so different from the Eulerian integral:
\be
F(x)=\int_0^{\infty} dt \frac{e^{-t}}{1+xt}
\ee
Which is known to be tractable with asymptotic methods \cite{Euler}. We can even resum the series using the Borel summation method\footnote{
Again, see Appendix A for details}
 to obtain:
\be
\label{Result1}
\Delta S^f = \Big(  \frac{G^{22}_{23} \big(\frac{1}{f^2 \epsilon^2}\big|^{1/2,1/2}_{0,1,-1/2} \big)}{2 \sqrt{\pi}}   \Big)  \Delta S^{\infty}
\ee
Where $\Delta S^{\infty}$ is the same as that in \ref{Minkfinty}.
 Although there is no proof ensuring resummation is unique, we present this as the correct solution for the entanglement entropy as a function of $f$.
 It is monotonic as expected and has the correct asymptotic expansion.
 One can also check that it \emph{isn't analytic at the origin} as expected. So the appearance of the tachyon, and therefore the absence of stability,
for $f < 0$ appears as non-analyticity of $\Delta S^f$.

\begin{figure}[h!]
\label{MinkowskiDS}
\includegraphics[width=\textwidth]{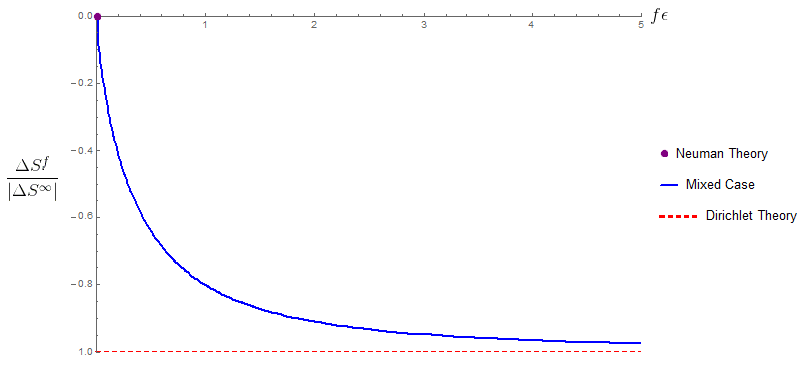}
\centering
\caption{This plot shows $\Delta S^f/|\Delta S^{\infty}|$ vs. $f \epsilon$ for $D=4$ and $m^2=0$. Notice it is decreases monotonically}
\end{figure}

One can now plot the whole interpolation between theories, see figure \ref{MinkowskiDS}.
 Notice $f$ only appears in the combination $f \epsilon$, which encodes the ratio of the renormalization scale to the cutoff.

Notice also that it depends not on the area of the horizon, but the area of the intersection of it with the boundary. So it's \emph{subleading} to the usual area law. This plus its monotonicity is reminiscent of the $g$-theorem for $D=2$, where we have:
\be
S=\frac{c}{6} \log \Big(\frac{\Lambda}{\epsilon} \Big)+\log(g)+c_0
\ee
The first term takes the place of the area law ($c$ is the central charge),
 the second term is a constant (since $n=0$ here) which depends on the boundary physics,
 and the last term is a constant which doesn't.
 The $g$ term has been proven to decrease monotonically, and this has been generalized to higher dimensions more recently \cite{Casini:2018nym}.
 Our result are illustrative of that this generalization, again building on \cite{Berthiere:2016ott}.

Since our results are in $D=4$, we should comment specifically on the ``Boundary F-theorem'' conjecture \cite{Estes:2014hka} \cite{Gaiotto:2014gha} \cite{Kobayashi:2018lil}.
The conjecture, which is a generalization of the $g$-theorem specific to $D=4$, states that for a 4D CFT with a boundary RG flow, the constant term in the entanglement entropy of a hemispherical region centered on the boundary, with bulk contribution subtracted, is monotonically decreasing along the flow.
Although we have found the entanglement for a planar region, we may still assess the behavior of any constant term for sake of comparison.\footnote{We thank the reviewer for recommending this comparison}
We see from \ref{Result1} and \ref{Minkfinty} that our results are entirely of the form of a perimeter law with \emph{zero} constant term, since:
\be
\Delta S^f \sim \Delta S_{\infty} \sim \frac{\Lambda}{\epsilon}
\ee
Therefore the monotonicity of the constant term is satisfied, but trivially so being always zero. Our $D=4$ monotonicity result for the coefficient of the perimeter term is therefore more comparable to \cite{Casini:2016udt}.

\section{Anti-de Sitter Space}

\label{SAdS}

Now we will turn to the free scalar (\ref{ScalarFieldTheory}) in Anti-de Sitter space, with (Euclidianized) metric:
\be
ds^2=\frac{L^2}{z^2}\big(dz^2+d \tau^2+d \vec{x}\cdot d \vec{x} \big) \qquad z \geq 0
\ee
Topologically this is the same as the previous section, and we will again use $x_1=0$ to define the Rindler splitting. For the theory 
(\ref{ScalarFieldTheory}) in an Anti-de Sitter background, it is convenient to introduce the quantity:
\be
\label{nudef}
\nu=\sqrt{d^2/4+m^2L^2} \qquad \Delta_{\pm}=\frac{d}{2}\pm \nu
\ee
For example, the small $z$ limiting behavior for any solution to the equation of motion goes as:
\be
\label{zexp}
\phi(z)=p_1 z^{\Delta_-}+\ldots+p_2z^{\Delta_+}
\ee
And for the range of masses given by: 
\be
0\leq\nu\leq1
\ee
A one parameter family of boundary conditions is permissible just as in the Minkowski case: 
\be
p_2=f p_1 \qquad f \ge0
\ee
Where we see the mass dimension of f is [$f$]$=2 \nu$.\footnote{Because we will be integrating over $\nu$ at various points,
 it will be necessary to be careful about this implicit $\nu$ dependence in $f$}
Just as before, as long as $f > 0$ the theory is well defined (as was shown in \cite{Ishibashi:2004wx}).
The $f=0$ case is traditionally still called in ``Neuman Theory'' and the $f \to \infty$ the ``Dirichlet Theory''.
 As was pointed out in \cite{Gubser:2002zh}, it is important to note that in the limit $\nu \to 0$
 the spectra for different $f$ all degenerate and so all give rise to the same quantum theory.\footnote{Nevertheless, 
there is still a 1 parameter family of possibilities due to the appearance of an additional term in the expansion (\ref{zexp})
 for $\nu=0$. This is just as with degeneracy for ordinary differential equations wherein an ``additional solution'' appears.
 One could consider this family additionally but we will not pursue this here}

As mentioned in Section (\ref{SI}), in the Dual Conformal Field Theory 
the $f$ parameter acts as a coupling constant for a double trace deformation of the Neuman theory,
 and this deformation generates a Renormalization group flow between a theory with operators of scaling dimension $\Delta_-$ to one of $\Delta_+$.

Unlike with the Minkowski case it is this Conformal Field Theory Dual we have in mind,
 and we are thinking of our results as a holographic calculation of the (Rindler) entropy in the Dual theory.
 We are able to relate the two using the proposal (\ref{FLMp}). In our case, we are only interested in the entropy difference, so we have:
\be
\label{dFLM}
\Delta S_{CFT}^f=\underbrace{\frac{\Delta \delta A^f}{4G}}_{geometric}+\underbrace{S_{AdS}^f}_{entropic}
\ee
We obtain (\ref{dFLM}) by noting that the classical contributions cancel (because the classical solution for the theories we consider are $\phi=0)$),
 that we have no curvature couplings (which would give $S_{Wald}$ in (\ref{FLMp})),
 and by assuming that any renormalization counterterms do not depend on $f$.\footnote{This
 is a reasonable assumption but it is not guaranteed since in principle there may be finite boundary counterterms associated with $f$. We will neglect this possibility here.}
This is all precisely the same as in \cite{Miyagawa:2015sql} and \cite{Sugishita:2016iel}
 where the $f=0,\infty$ cases were computed.
 We seek to extend their calculation to general $f$ and compare results to expectations based on the irreversibility of the Renormalization Group (\ref{SSA_Req}) as a way of checking the consistency of the proposal (\ref{FLMp}). So we must compute two terms,
 an \emph{entropic} contribution and a \emph{geometric} contribution.
Both can be obtained from the Greens function.

The Greens function is most easily found not by spectral theory but by solving:
\be
\hat{D} G(x,x')=\frac{1}{\sqrt{g}} \delta(x-x')
\ee
The metric factor is chosen so that: 
\be
\int dV f(x)\Big((\Box-m^2) G(x,x') \Big) =f(x')
\ee
The solution requires nothing other than standard Sturm-Liouville techniques and resembles finding the classical static electric field Greens function in cylindrical coordinates as in \cite{Jackson}. This task was first accomplished for general $f$ in \cite{Gubser:2002zh}. We have:
\be
\label{AdSG}
\Delta G^f=\frac{-\sin(\pi \nu) L }{\pi} \int d \tilde{k} \alpha_k^f (z z')^{d/2}K_{\nu}(k z)K_{\nu}(k z')e^{i k |\Delta \vec{ r}| \cos(\theta)}
\ee
With: 
\be
\alpha_k^f=\frac{2^{2 \nu}f \Gamma(1+\nu)}{k^{2 \nu} \Gamma(1-\nu)+2^{2 \nu}f \Gamma(1+\nu)}=\Big(\frac{k^{2 \nu} \Gamma(1-\nu)}{2^{2 \nu}f \Gamma(1+\nu)}+1 \Big)^{-1}
\ee
And: 
\be
d \tilde{k}= \frac{\Omega_n k^{d-1}\sin(\theta)^n dk d\theta}{(2 \pi L)^d}
\ee
And where $\Delta \vec{r}$ refers to the boundary directions only.

\subsection{Entropic Contribution}
\label{entropiccont}

\begin{figure}[h!]
\label{AdSEntropyPlot}
\includegraphics[width=\textwidth]{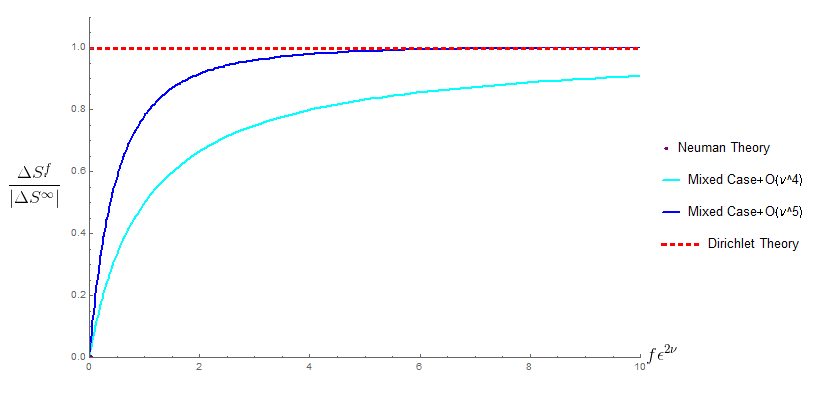}
\centering
\caption{This plot shows $\Delta S^f_{AdS}/|\Delta S^{\infty}_{AdS}|$ vs. $f \epsilon$ for $\nu=\frac{1}{2}$ for any $D = 5$ .
 Notice it is increases monotonically}
\end{figure}

We may proceed to get the bulk entanglement entropy just as in Section \ref{SM}, this time by expanding\footnote{And 
the same theorem guarantees we will get at least an asymptotic series}:
\be
\label{AdSexp}
\alpha_k^f=\sum_i\Big(-\frac{k^{2 \nu} \Gamma(1-\nu)}{2^{2 \nu}f \Gamma(1+\nu)}\Big)^{i}
\ee
For full details, see Appendix B. As it turns out there is a complication which is that for the Anti-de Sitter the analogue of (\ref{Seq2}):
\be
W=\frac{1}{2} \int_0^{\nu} d\nu^2 TrG
\ee
Where we may integrate from $\nu=0$ since $\Delta W_{\nu=0}=0$ as explained above. The issue is \emph{this} integral will be intractable.
 However if we expand in $\nu$, which is reasonable since $0\le \nu \le 1$,
 we will be able to proceed and will even be able to resum the series (\ref{AdSexp}) in $f$ order by order in $\nu$
\emph{without} asymptotic methods. The result, which does not converge,
 can be interpreted as an asymptotic expansion in small $\nu$ For example, for $D=5$ we obtain:
\be
\label{Result2b}
\Delta S^f=\Delta S^{\infty} \Big(\frac{f \epsilon^{2 \nu}}{1+f \epsilon^{2 \nu}} \Big)+s_{4,1} \Big(\frac{f \epsilon^{2 \nu}}{(1+f \epsilon^{2 \nu})^2} \Big)+\mathcal{O}(\nu^5)
\ee
Where:
\be
\Delta S^{\infty}=\frac{1}{72 \pi} \Big(\frac{\Lambda^2}{\epsilon^2} \Big) \nu^3
\ee
Which agrees with \cite{Miyagawa:2015sql} and \cite{Sugishita:2016iel}, and where:
\be
\label{Sbulk5}
s_{4,1}=\frac{1+\gamma_e+\log(4)+\psi^0(3/2)}{96 \pi} \Big(\frac{\Lambda^2}{\epsilon^2} \Big) \nu^4
\ee
This is monotonic, just as in the Minkowski case, however here it is monotonically \emph{ increasing}. See Figure \ref{AdSEntropyPlot}

It is possible to compute the additional terms systematically, for any $D \ge 4$.\footnote{For
 $D=3$ there are additional divergences which prevent the integral expressions from being tractable even after expansion,
 see \cite{Miyagawa:2015sql}}.
 For all cases the results have the structure:
\be
\label{Sbulkany}
\Delta S_{AdS}^f=\sum_{i=3}^{\infty} \sum_{j=0}^{i-3} s_{ij} \nu^i \phi(-\frac{1}{f \epsilon^{2 \nu}}, -j,0)
\ee
Where $\phi$ is the Hurwitz Lerch $\phi$ function.
 Only terms with $j=0$ will be nonzero in the limit $f \to\infty$ and these will only contribute for odd $i$ for $3 \leq i \le D-2$.
 For $D=5$ this is only the $\nu^3$ term which is why we were able to write it in the form (\ref{Sbulk5}).

The expression we provided is written in terms of the dimensionless parameter $f \epsilon^{2 \nu}$, so we have implicitly had in mind fixing this quantity,
 expanding in it, and then resumming order by order in $\nu$. 
This form is useful for considering e.g. varying $f$ with $\epsilon$ fixed,
 which interpolates between Neuman and Dirichlet theories.
 However for some fixed $f$ we would like to take $\epsilon \to \infty$ since it is an ultraviolet regulator in the conformal field theory
 and we are really only interested in finite or divergent terms in this limit.
 Which terms survive will depend on the choice of $D$ and $\nu$, another reason (\ref{Sbulk5}) and (\ref{Sbulkany}) are useful.
 For the case $D=5$ and $\nu=\frac{1}{2}$ again, we get for example\footnote{We
 could not provide a similar expression in the Minkowski case because the expression was not analytic for small $f \epsilon$}
:
\be
\Delta S^f_{AdS}= \frac{3(\gamma_e+\log(4)+\psi^0(3/2))+11}{4608 \pi} ( \frac{f \Lambda^2}{\epsilon})-\frac{3(\gamma_e+\log(4)+\psi^0(3/2))+7}{2304 \pi} (f^2 \Lambda^2) +\mathcal{O}(\nu^5)
\ee
Note the second term is a finite contribution, while both are proportional to the area in the dual conformal field theory.

It is worth noting that as a bonus we may obtain the zero temperature Free Energy (The Euclidean connected function), which is related to the vacuum energy sought in \cite{Gubser:2002zh}.
 In \cite{Gubser:2002zh} the authors noted monotonicity of this quantity based on the integral expression \ref{AdSG} but did not compute the integral.
 We may however proceed with the same strategy as for the entropy to compute, for example:
\be
\label{FrEnergy}
\Delta W^f=- \Big(\frac{f \epsilon^{2 \nu}}{1+f \epsilon^{2 \nu}} \Big) \Big(\frac{\Lambda^d}{\epsilon^d} \Big) \Big(\frac{\Omega_n}{2^{d+1} \pi^{d-1}} \Big)  \frac{ \Gamma(\frac{d-1}{2}) \Gamma(\frac{d}{2})^2}{3 d \Gamma(\frac{d+1}{2})}\nu^3+\mathcal{O}(\nu^4)
\ee
Which in e.g. $D=5$ gives: 
\be
\Delta W^f=-\frac{1}{144 \pi^2} \Big(\frac{f \epsilon^{2 \nu}}{1+f \epsilon^{2 \nu}} \Big) \Big(\frac{\Lambda^4}{\epsilon^4} \Big) \nu^3 +\mathcal{O}(\nu^4)
\ee
We note that $\frac{\Lambda^4}{\epsilon^3}$ is just the volume of the conformal field theory, so the coefficient of this may be
interpreted as the free energy density (at zero temperature).
All the higher order terms can be computed systematically as described previously. Notice the monotonicity appears as expected.

\subsection{Geometric Contribution}

In order to make a prediction for the entanglement entropy of the dual conformal field theory using the proposal (\ref{FLMp}),
 we must also compute the area term which comes from 1-loop stress tensor backreaction on the geometry.
In general, this could result in a different Ryu-Takayangi minimal surface in the bulk, but since we have chosen the horizon to be defined by $x_1=0$ the unbroken symmetries ensures this will not change in our case and we need only compute the shift in area for this same surface.

We must keep in mind also that we should use the same expansion employed when computing the entropic contribution.
Namely, we fix a $z$ cutoff for the space and expand in the $\nu$ independant quantity $f \epsilon^{2 \nu}$. 
This means we will need to be solving the (linearized) Einstein equations with appropriate boundary conditions imposed at $z=\epsilon$\footnote{Equivalently,
 we may think of choosing boundary conditions by requiring that in the $\epsilon \to 0$ limit we have the same boundary as the Neuman theory since this fixes the ultraviolet fixed point}

We proceed as follows. First, we must obtain the stress tensor. We may get this by ``point splitting''
the Greens function (see e.g.  \cite{Moretti:1998rs}). That is we define:
\be
\label{ptsplit}
 \langle T_{ab} \rangle =\lim_{x'\to x}\big( \langle \hat{T}_{ab}(x,x') \rangle -Z_{ab}\big)+g_{ab}Q
\ee
Where the first term is the classical stress tensor promoted to a quantum operator, but ``point split'' so it isn't evaluated at coincident points.
 The second term is a quantity meant to remove the divergences from the first quantity.
It can be rather difficult to determine but depends only on geometric invarients. The final term $Q$ is meant to enforce:
\be
\nabla^a \langle T_{ab} \rangle =0
\ee
For us, the issue is simplified since we are only interested in the \emph{difference} in stress tensors for different boundary conditions.
 The divergent part $Z_{ab}$ is removed from this automatically.\footnote{In 
principle it is possible that some finite part remains, since we are comparing two different theories,
 not two states in the same theory (for which the subtraction is guaranteed to be correct by a theorem \cite{WaldQFT}).
 We will neglect this possibility since we appear to obtain the correct answer when our results reduce to known results obtained by other methods.}

The classical (Hilbert) stress tensor is:
\be
T_{ab} =\frac{2}{\sqrt{g}} \frac{\delta (\sqrt{g}\mathcal{L})}{\delta g^{ab}}
=
\qquad =2 \frac{\delta \mathcal{L}}{\delta g^{ab}}+g_{ab} \mathcal{L}
\ee
Which for our case is:
\be
T_{ab}= \partial_a \phi \partial_b \phi+\frac{1}{2} g_{ab}(g^{cd}\partial_c \phi \partial_d \phi+m^2 \phi^2)
\ee
Where recall $m^2$ is related to $\nu$ by (\ref{nudef}) which is important for a $\nu$ expansion. The promotion of $T_{ab}$ to a point-split operator and taking its expectation value can be accomplished by replacing the terms in (\ref{ptsplit}) with the Greens function or the appropriate derivative.
See Appendix B for the details of this and the rest of the calculation.

So using (\ref{ptsplit}) we may obtain an expansion for difference in stress tensors. To lowest order in $\nu$ we get:
\be
\label{stresstensor}
 \langle T_{ab} \rangle =-\big(\frac{f \epsilon^{2 \nu}}{1+f \epsilon^{2 \nu }} \big) \big( \frac{\Omega_n d^2 \Gamma(\frac{d-1}{2}) \Gamma(\frac{d}{2})^2 }{2^{d+4} \Gamma(\frac{d+3}{2}) \pi^{d-1}} \big) \frac{\nu}{L^D} g_{ab}+\mathcal{O}(\nu^2)
\ee
Where $\Omega_n$ is the area of the $n$-sphere. For $D=5$ this is:
\be
 \langle T_{ab} \rangle =-\frac{f \epsilon^{2 \nu}}{1+f \epsilon^{2 \nu}}(\frac{\nu}{15 L^5 \pi^2})g_{ab}+\mathcal{O}(\nu^2)
\ee
 In order to obtain (\ref{stresstensor}), one must expand as in the previous subsection. This means cutting off the space at $z=\epsilon$,
 (\ref{stresstensor}) should therefore be thought of as the boundary value of the stress tensor which interpolates between Neuman and Dirichlet as a function of $z$. The contribution goes to zero as $\epsilon \to 0$ for fixed $f$ as it should, since all theories approach the Neuman fixed point
 in the ultraviolet.

\begin{figure}[h!]
\label{plotAdS}
\includegraphics[width=\textwidth]{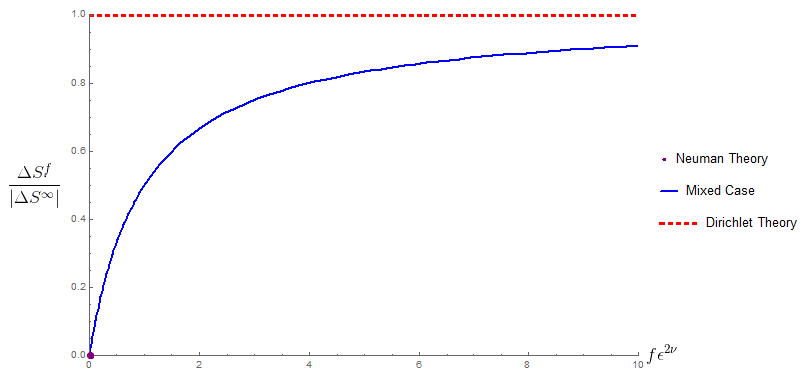}
\centering
\caption{This plot shows $\Delta S^f_{CFT}/|\Delta S^{\infty}_{CFT}|$ vs. $f \epsilon$ to lowest order in $\nu$ for any $D \ge 4$ .
 Notice it is increases monotonically.}
\end{figure}

Because this contribution is proportional to the metric, at this order in $\nu$ the backreaction is equivalent to a simple shift of the cosmological constant:
\be
 \langle T_{ab} \rangle =\lambda g_{ab} \to \delta \Lambda_{c.c.}=-8 \pi G \lambda
\ee
 The cosmological constant is related to the $AdS$ radius $L$ by:
\be
\Lambda_{c.c.}=-\frac{(d)(d-1)}{2 L^2}
\ee
So:
\be
\delta L = -\frac{2 \sqrt{2} G L^3 \pi \delta \lambda}{d(d-1)}+\mathcal{O}(G^2)
\ee
Meanwhile the classical result for the area is:
\be
A=\Lambda^n\int_{\epsilon}^{\infty} (L/z)^{n+1}=\frac{L^{n+1}}{n} \frac{\Lambda^n}{\epsilon^n}
\ee
Putting this altogether we get:
\be
\frac{\Delta \delta A^f}{4G}=-\frac{\Lambda^n}{\epsilon^n} \frac{2 \pi L^{3+n}}{d(d-2)} \Delta\lambda^f
\ee
And so finally we have:
\be
\label{AdSAterm}
\frac{\Delta \delta A^f}{4G}+\mathcal{O}(\nu^2)=
\big(\frac{f \epsilon^{2 \nu}}{1+f \epsilon^{2 \nu }} \big)
\big( \frac{\Omega_n d^2 \Gamma(\frac{d-1}{2}) \Gamma(\frac{d}{2})^2 }{2^{d+3} d(d-2) \Gamma(\frac{d+3}{2}) \pi^{n}} \big)
 \frac{\Lambda^n}{\epsilon^n}  \nu
+\mathcal{O}(\nu^2)
\ee
Which for $D=5$ again for example is:
\be
\frac{\Delta \delta A^f}{4G}=\frac{f \epsilon^{2 \nu}}{1+f \epsilon^{2 \nu}}\frac{\Lambda^2}{\epsilon^2}\frac{\nu}{60 \pi}+\mathcal{O}(\nu^2)
\ee
Note that this is a monotonic contribution to the boundary area law. It agrees with \cite{Miyagawa:2015sql} and \cite{Sugishita:2016iel}
 when $f \to \infty$. Since it is  of order $\nu^1$ whereas (\ref{Sbulkany}) was of order $\nu^3$ it is always the \emph{leading} contribution.
 So the geometric contribution leads the entropic contribution \emph{for any} $f$ (for Rindler space).
 So (\ref{AdSAterm}) is our prediction for the dual field theory Rindler entropy to lowest order in $\nu$. That is, to emphasize:
\be
\label{Result2}
\Delta S_{CFT}^f=
\frac{\Delta \delta A^f}{4G}+\mathcal{O}(\nu^2)=
\big(\frac{f \epsilon^{2 \nu}}{1+f \epsilon^{2 \nu }} \big)
\big( \frac{\Omega_n d^2 \Gamma(\frac{d-1}{2}) \Gamma(\frac{d}{2})^2 }{2^{d+3} d(d-2) \Gamma(\frac{d+3}{2}) \pi^{n}} \big)
 \frac{\Lambda^n}{\epsilon^n}  \nu
+\mathcal{O}(\nu^2)
\ee
We can plot this as in Section (\ref{SM}). See Figure \ref{plotAdS}
Higher order contributions can be computed systematically by proceeding with point splitting and solving the bulk Einstein Equations.
 See Appendix B for full details. The results have a structure similar to (\ref{Sbulkany}).

We can of course take the $\epsilon \to 0$ limit for some choice of $\nu$ just like in the previous section, and the results are similar.

\subsection{Irreversability}

The monotonicity of the interpolation of the entropy of the conformal field theory found in the previous section
 is reminiscent of the behavior of ``c-functions'' which capture the irreversibility of the renormalization group flow.
  On the other hand our entropy apparently \emph{increases} rather than decreases which is the opposite of what we would expect.

In order to clarify this issue we need to compare with explicit results on the irreversibility of the renormalization group flow
 and its relation to entanglement entropy. The most general result to date is the theorem in \cite{Casini:2017vbe} which was mentioned in the introduction.
 To be more explicit, let $\Delta S(r)$ be the difference in entanglement entropy between the deformed theory and the theory at the UV fixed point
 (just as in (\ref{entropydiff})) for a \emph{spherical} region of radius $r$. We can define the following quantity:
\be
\tilde{\mu}(r)=\frac{S ' (r)}{(d-2)r^{d-3}}
\ee
Where here we are thinking of $d$ as the spacetime dimension as in the conformal field theory dual to Anti-de Sitter space of dimension $D=d+1$.
 We can think of this as the coefficient of the area term. The result of \cite{Casini:2017vbe} says that this $\mu(r)$ acts as a c-function in any dimension:
\be
\label{CasiniCond}
\tilde{\mu}'(r) \le 0
\ee
That is it is monotonically decreasing as we go from ultraviolet to infrared with increasing spherical radius.

If we think of our Rindler result as representing that of a sphere of infinite radius $r \sim \Lambda$ the result in the previous section satisfies (\ref{CasiniCond}) trivially:
\be
\tilde{\mu}'(r)=0
\ee
This is because for any value of $f$ the Rindler Entropy is in the deep infrared regime.
 In order to compare with irreversibility expectations nontrivially, we may do one of two things:

1. Because our result to lowest order comes entirely from the shift in the cosmological constant, it would seem like we could extend our results
 to lowest order in $\nu$ to a spherical entangling surface using \cite{Ryu:2006bv}.
 If we do this naively, the inequality (\ref{CasiniCond}) will actually be violated.
 This is because, as was pointed out in \cite{Sugishita:2016iel}, 
there is a $\nu^1$ term that appears in the entropic contribution for finite $r$ which exactly cancels the geometric contribution,
 at least for the $f=\infty$ case. The important term is of order $\nu^3$.
 This resolves the puzzle pointed out in \cite{Miyagawa:2015sql}
 regarding consistency of higher order $\nu$ terms with expectations based on the $a$-theorem,
 but also means we cannot trivially extend our results to a spherical surfaces to the necessary order of $\nu$.
 This is because past first order the backreaction on the geometry is not reducible to a shift of the cosmological constant and we cannot e.g. guarantee
 the minimal surface remains the same.
 Nevertheless, if the interpolation found for the Rindler results extends to spherical entangling surfaces anyway,
 which appears likely since it appears to be inherited almost directly from the form of the Greens function,
 then the proof of the consistency of the proposal (\ref{FLMp}) with (\ref{CasiniCond}) in the $f \to \infty$ case found in \cite{Sugishita:2016iel}
 extends immediately
 to all $f$. So our results are highly suggestive, but \emph{not} a proof, of the consistency at finite $f$.

2. In $d=2$ dimensions ($AdS_3$) the area term is proportional to the central charge, so the interpolation would be directly interpretable.
 However we were not able to obtain the entropic contribution in this case. For $f \to \infty$ it was found that a naïve extension of the procedure in
 Section \ref{entropiccont} gave the correct result.
 If this holds true for finite $f$ then the interpolation \ref{Sbulkany} will remain true and the consistency found in \cite{Sugishita:2016iel}
 will extend to finite $f$.

It is also worth noting that the free energy computation (\ref{FrEnergy}) is consistent with the $c$-theorem,
 but this was already known in \cite{Gubser:2002zh} \cite{Gubser:2002vv}.

So our results are strongly suggestive, but do not concretely prove, that the proposal \ref{FLMp} is consistent with irreversibility of the renormalization group for finite $f$.

\section{Discussion and Conclusions}

\label{SD}

In this paper, we computed the entanglement entropy for mixed boundary conditions in half Minkowski space and
 in the context of the AdS/CFT correspondence. 
In both cases, the result was a monotonic interpolation between conformal fixed points as a function of a dimensionless combination 
of the boundary coupling and the cutoff. 
In the case of half Minkowski space, our results build on earlier results and are illustrative of the generalization of the g-theorem to higher dimensions found in \cite{Casini:2018nym}.
 In Anti-de sitter space our results fill in the interpolation between Neuman and Dirichlet theories as already computed in \cite{Miyagawa:2015sql} and \cite{Sugishita:2016iel}
 and offer an opportunity for a consistency check of the proposal for $1/N$ corrections to the holographic entanglement entropy formula.

We have also commented on the non analyticity of entropy difference at $f=0$ which was observed in \cite{Berthiere:2016ott}, where it
was suggested that it may be indicative of a phase transition. Indeed, a tachyon appears precisely at this point. The theory would need to be embedded
in a larger theory to determine the new phase since our free theory simply becomes divergent.

It may be possible to extend the AdS/CFT result to higher order using the recent proposal\cite{Engelhardt:2014gca} \cite{Dong:2017xht}:
\be
S_{CFT}=ext[\frac{ \langle A \rangle }{4G}+S_{AdS}]
\ee
But it is not entirely clear how to make sense of the gravitational backreaction beyond 1-loop
 since gravity is not renormalizable and since the next order would inevitably involve quantum gravitational interactions with the matter field $\phi$
 (for us the backreaction involved classical gravity responding to the 1-loop stress tensor, which is a tadpole diagram).

It would be interesting to extend the results to higher spin. For spin 1, there is the additional possibility of topological terms which are supposedly probed by the entanglement entropy. Combined with mixed boundary conditions, there is a full SL(2,Z) space of theories 
(see \cite{Cottrell:2017gkb} for summary and elaboration on possibilities) 
which can be explored and it would be interesting to see how the entropy transforms under this group.
For spin 2, it has been suggested \cite{Cottrell:2018skz} that the mixed boundary conditions give rise to quantum gravity on the boundary. 
In this case it is not even clear what the analogue of the formula (\ref{FLMp}) would be, making it especially interesting though perhaps problematic.

As mentioned, the results confirm expectations based on the irreversibility of the renormalization group flow. As mentioned in \cite{Casini:2018nym} it would be desirable to extend these results as much as possible, for example to boundaries of different codimension.

The fact that the entropy difference (\ref{entropydiff}) is so easily computable in this example may make a comparison with the entropy bounds 
\cite{Casini:2008cr} \cite{Bousso:2014sda} interesting.

The Ryu-Takayangi formula has been used to derive the linearized Einstein equations in the bulk from
 the ``first law of entanglement entropy'' on the boundary (see e.g. \cite{Lashkari:2013koa}  \cite{Faulkner:2013ica} \cite{Faulkner:2017tkh})
 Extending these results to include quantum corrections is important and the example in this paper may provide an interesting test case
 for exploratory purposes.

The solubility of this model may be useful in general for exploring holographic renormalization group flows generated by $\frac{1}{N}$ suppressed effects.

\appendix
\section{Appendix A}
We choose $D=4$ with $m^2=0$. Since $m^2=0$ it is actually more convenient to work with the heat kernal. We have:
\be
K=-\sum_{i=0}^{\infty} \int_{-\infty}^{+\infty} \int_0^{\infty} \int_0^{\pi} d \kappa dk d\theta \frac{ k^2  \text{sin}(\theta )}{4 \pi ^3}\left(-\frac{k^2}{f^2}\right)^ie^{2 i z \kappa -s \left(k^2+m^2+\kappa ^2\right)+i k \Delta r \text{cos}(\theta )}
\ee
Where we have already expanded in large $f$. Since this expansion of the integrand is analytic the resulting integral, if finite term by term,
 givesnan asymptotic series.
 It is convenient to use polar coordinates for the boundary directions, in which case:
\be
\Delta r =| r| \sqrt{2(1 - \cos(w))}
\ee
We can then perform the partial trace:
\be
K_{w}=\int_0^{\infty} r dr K=\sum_{i=0}^{\infty} \frac{e^{-m^2 s-\frac{z^2}{s}} \left(-\frac{1}{f^2 s}\right)^i \Gamma\left(\frac{1}{2}+i\right)}{\pi ^{5/2} (-8 s+8 s \text{cos}(w))}
\ee
Since the entropy does not depend on terms linear in $n$ the Sommerfeld formula can be applied as:
\be
K_{n} \sim - \pi \text{Residue}[K_{w} \cot(\frac{w}{2n}), w=0]
\ee
Since this is the only pole. To get the entropy we then take:
\be
\big( (\partial_n-1) K_n \big)|_{n=1}
\ee
And integrate over $z$ and the boundary directions. We get:
\be
\partial_n K_s =\sum_{i=0}^{\infty} \Lambda \frac{\left(-\frac{1}{f^2 s}\right)^i \Gamma\left[\frac{1}{2}+i\right]}{24 \pi  s^{1/2}}
\ee
We integrate over $s$ to get the the entropy:
\be
\label{AoA1}
S=-\frac{1}{2} \int \frac{ ds}{s} \partial_n K_s=\sum_{i=0}^{\infty}S_i=\sum_{i=0}^{\infty}-\frac{\left(-\frac{1}{f^2 \epsilon^2}\right)^i  \Gamma\left[\frac{1}{2}+i\right]}{24 (\pi +2 i \pi )} \frac{\Lambda}{ \epsilon}
\ee
This sum as divergent, but it may be Borel resummed since the sum:
\be
\sum \frac{S_i}{i}=-\frac{f \Lambda \text{ArcSinh}\left(\frac{1}{f \epsilon }\right)}{24 \sqrt{\pi }}
\ee
is convergent. In Borel summation one takes the divergent sum $\sum A_i$, replaces it with $\sum A_i/i$, and integrates $\int_0^{\infty} dt e^{-t/z}$ where $t$
is the asymptotic expansion parameter, for us $t= f^2 \epsilon^2$ since this was what appeared in (\ref{AoA1}). This gives the result:
\be
\Delta S^{f}=\Delta S^{\infty}\frac{\text{MeijerG}\left[\left\{\left\{\frac{1}{2},\frac{1}{2}\right\},\{\}\right\},\left\{\{0,1\},\left\{-\frac{1}{2}\right\}\right\},\frac{1}{f^2 \epsilon^2}\right]}{2 \sqrt{\pi }}
\ee
As reported in Section \ref{SM}.

\section{Appendix B}

\subsection{Entropic Contribution}

We will proceed for $D=5$ since it is difficult to obtain a general expression in terms of $D$ even as each $D \ge 4$ appears tractable case by case. If
we expand the Green's function in large $f$ we may perform the $k$ integral. We get:
\be
\sum_{i=0}^{\infty}-\frac{2^{-6-2 i \nu } z^{-2 i \nu } (4 \pi) \left(-\frac{\Gamma(1-\nu )}{f \Gamma[1+\nu ]}\right)^i \Gamma[2+(-1+i) \nu ] \Gamma[2+i \nu ]
   \Gamma[2+\nu +i \nu ] \text{sin}(\pi  \nu )}{L^3 \pi ^{7/2}} \times 
\ee
\be
\text{HypergeometricPFQRegularized}\left[(2+i \nu ,2+(-1+i) \nu ,2+\nu +i \nu ),(2,\frac{5}{2}+i \nu ),-\frac{\Delta r^2}{4
   z^2}\right] 
\ee
One proceeds from here exactly as in Appendix A, however one integrates over $\nu$ instead of the heat kernal parameter $s$. In order to perform
this integral one must expand in small $\nu$. One finds terms:
\be
\sim i^k(-\frac{1}{f})^i
\ee
Where $k$ is less than the order of $\nu$ minus 3. These may be resummed over $f$ term by term, because they are analytic power series in $f$.
 This is what gives rise to the $\phi$ functions and the structure
(\ref{Sbulkany}).
However the series is not convergent when summed over $\nu$ and we should recall that this is only an asymptotic series.

\subsection{Geometric Contribution}
We may find the bulk stress tensor by expanding the integrand of Green's function in large $f$,
 taking appropriate derivatives, and taking the coincidence limits. We obtain:
\begin{multline}
\Delta  \langle \phi \phi \rangle ^f= \\
\sum_{i=0}^{\infty} -\frac{    \Omega_n \Gamma (\frac{1}{2} (-1+d)) \left(-\frac{\Gamma(1-\nu )}{f \Gamma(1+\nu
   )}\right)^i \Gamma\left(\frac{d}{2}+(-1+i) \nu \right) \Gamma\left(\frac{d}{2}+i \nu \right) \Gamma\left(\frac{d}{2}+\nu +i \nu \right) \text{sin}(\pi  \nu)
   }{2^{+1+d+2 i \nu } \pi ^{d} L^{d-1} z^{2 i \nu } \Gamma\left(\frac{d}{2}\right) \Gamma\left(\frac{1}{2} (1+d+2 i \nu )\right)}
\end{multline}
\begin{multline}
\Delta  \langle  \partial_z \phi \partial_z \phi \rangle ^f= \\
\sum_{i=0}^{\infty}- \left(d^2+4 d i \nu  (1+i \nu )+4 \nu ^2 \left(-1+2 i^2 (1+i \nu )\right)\right) \times \\
\frac{ \Omega_n
   \Gamma\left(\frac{1}{2} (-1+d)\right) \left(-\frac{\Gamma(1-\nu )}{f \Gamma(1+\nu )}\right)^i \Gamma\left(\frac{d}{2}+(-1+i) \nu \right)
   \Gamma\left(\frac{d}{2}+i \nu \right) \Gamma\left(\frac{d}{2}+\nu +i \nu \right) \text{sin}(\pi  \nu )}{ 2^{4+d+2 i \nu } L^{d-1} \pi ^{d} z^{2+2 i \nu } \Gamma\left(\frac{d}{2}\right)
   \Gamma\left(\frac{1}{2} (3+d+2 i \nu )\right)}
\end{multline}
\begin{multline}
\Delta  \langle  \partial_x \phi \partial_x \phi \rangle ^f= \\
\sum_{i=0}^{\infty}-\frac{ \Omega_n \Gamma\left(\frac{1}{2} (-1+d)\right) \left(-\frac{\Gamma(1-\nu )}{f \Gamma(1+\nu
   )}\right)^i \Gamma\left(1+\frac{d}{2}+(-1+i) \nu \right) \Gamma\left(1+\frac{d}{2}+i \nu \right) \Gamma\left(1+\frac{d}{2}+\nu +i \nu \right) \text{sin}(\pi  \nu
   )}{2^{2+d+2 i \nu } L^{d-} \pi ^{d} z^{2+2 i \nu }\Gamma\left(1+\frac{d}{2}\right) \Gamma\left(\frac{1}{2} (3+d+2 i \nu )\right)}
\end{multline}
Where $x$ is any boundary coordinate (all the same due to boundary Lorentz symmetry and Euclideanization).
These are badly divergent, not even Borel summable, but they can still be treated as asymptotic series.
 We then have:
\begin{multline}
\Delta  \langle T_{xx} \rangle ^f =
 \frac{d+2}{2}\Delta  \langle  \partial_x \phi \partial_x \phi \rangle ^f + (1/2) \Delta  \langle  \partial_z \phi \partial_z \phi \rangle ^f + (1/2) (1/z^2) (\nu^2 - \frac{d^2}{4}) \Delta  \langle \phi \phi \rangle ^f
\end{multline}
\begin{multline}
\Delta  \langle T_{zz} \rangle ^f =
 \frac{d}{2}\Delta  \langle  \partial_x \phi \partial_x \phi \rangle ^f + (3/2) \Delta  \langle  \partial_z \phi \partial_z \phi \rangle ^f +  (1/2) (1/z^2) (\nu^2 - \frac{d^2}{4}) \Delta  \langle \phi \phi \rangle ^f
\end{multline}
One can then expand in $\nu$. One finds that:
\be
\Delta  \langle T_{xx} \rangle ^f-\Delta  \langle T_{zz} \rangle ^f=0+\mathcal{O}(\nu^2)
\ee
Further the $z$ dependence is $\sim z^{-2}$ so this first contribution is proportional to the metric.
 This means $\Delta Q^f=0$ automatically at this order and allows a quick determination of the leading order contribution to the shift in the area
 as explained in Section \ref{SAdS}.
 For higher order in $\nu$ there will be both further contributions proportional to the metric and ones with additional $z$ dependence.
 The former can be handled as in Section \ref{SAdS} but for the latter we must explicitly solve the Einstein equation.
 It actually appears that $\Delta Q^f$ remains 0, at least for the first few orders in $D=5$. The condition is:
\be
\nabla^a  \langle T_{ab} \rangle =(d \langle T_{xx} \rangle -n  \langle T_{zz} \rangle + z \partial_z  \langle T_{zz} \rangle ) \frac{z}{L^2}=0
\ee
A convenient metric ansatz is given by:
\be
ds^2=(\frac{L^2}{z^2})(dz^2+e^{2 h(z)} d \vec{x} \cdot d \vec{x})
\ee
The $zz$ component of the Einstein Equation then gives:
\be
-\frac{d(d-1) z h'(z)}{L^2}=T^z_z
\ee
Which can be integrated immediately, with a boundary condition set so as to recover $h(\epsilon) \to 0$ as $\epsilon \to 0$.
Then one simply needs to integrate:
\be
\frac{\Delta \delta A}{4 G}^f= \frac{(d-2) \Lambda^n }{4 G} \int_{\epsilon}^{\infty} dz (\frac{L}{z})^{d-1}\Delta h(z)
\ee
This can be done systematically but is very messy. As with the other contributions one may expand in $\nu$ and resum order by order in $f$.
For example in $D=5$ we can get the order $\nu^2$ term this way:
\be
\frac{\Delta \delta A}{4 G}^f|_{o(\nu^2)}=-\frac{f \epsilon^{2 \nu}}{(1+f \epsilon^{2 \nu})^2}\frac{73}{5400 \pi } \frac{\Lambda^2}{\epsilon^2} \nu^2
\ee

\acknowledgments

The author would like to thank Aki Hashimoto for many extensive discussions.


\end{document}